%% file: manuscript.tex
\documentclass[prl,reprint,superscriptaddress,amsmath,amssymb,floatfix, showkeys]{revtex4-2}

\usepackage{graphicx}
\graphicspath{{figures/}}
\usepackage{dcolumn}
\usepackage{bm}
\usepackage[utf8]{inputenc}
\usepackage[T1]{fontenc}
\usepackage[dvipsnames]{xcolor}
\usepackage{xspace}
\usepackage{ulem}

\newcommand{\jpsi}{J/\ensuremath{\psi}\xspace}
\newcommand{\jpsiexp}{\jpsi--007\xspace}
\newcommand{\Agt}{\ensuremath{\mathcal{A}_g(t)}\xspace}

\newcommand{\Cgt}{\ensuremath{\mathcal{C}_g(t)}\xspace}

\newcommand{\Cqgt}{\ensuremath{\mathcal{C}_{q/g}(t)}\xspace}
\newcommand{\Cbargt}{\ensuremath{\bar{\mathcal{C}}_g(t)}\xspace}
\newcommand{\Cbarqt}{\ensuremath{\bar{\mathcal{C}}_q(t)}\xspace}

\newcommand{\new}[1]{#1\xspace}

\begin{document}

\preprint{XYZ}

\title{Near-Threshold \jpsi $\to \mu^+\mu^-$ Photoproduction and \new{the Gluonic Gravitational Form Factors of the Proton}}

\input{authors}


\keywords{photoproduction, charm, mass, gravitational form factors, EMT form factors, proton structure}

\begin{abstract}
We report on the measurement of the two-dimensional differential cross section for near-threshold \jpsi~$\to \mu^+\mu^-$ photoproduction from the \jpsiexp experiment in Hall C at Jefferson Lab.
Our results agree with the previously published \jpsi~$\to e^+e^-$ results.
We extract the integrated photoproduction cross section and find no evidence for open-charm contributions.
A combined analysis of both decay channels \new{following a Holographic QCD approach yields improved experimental constraints} on the gluonic gravitational form factor \Cgt. 
Our results agree with recent lattice QCD calculations
\new{and we obtain \Cgt with a comparable statistical precision to lattice QCD}.
\new{Our results support a spatial picture where gluons dominate at larger radii with a confining inward pressure}. 
This work provides new input for exploring the mechanical properties of gluons inside the proton.
\end{abstract}

\maketitle

\section{Introduction}

The proton's mass and internal structure are predominantly determined by the dynamics of gluons, which are described in the framework of Quantum Chromodynamics (QCD). Recently, the measurement of near-threshold \jpsi photoproduction has provided a unique new tool for probing the gluonic gravitational form factors (GFFs)~\cite{Kharzeev:2021qkd,Mamo:2021krl,*Mamo:2022eui,Guo:2023pqw}. \new{GFFs parametrize the matrix elements of the QCD energy-momentum tensor, encoding} information about the mechanical properties of the proton, \new{such as the spatial density of its mass and internal forces} ~\cite{Teryaev:2016edw,Lorce:2018egm}.

Previous measurements from the \jpsiexp experiment demonstrated experimental constraints on the gluonic GFFs using the electron decay channel (\jpsi~$\to e^+e^-$)~\cite{Duran:2022xag}. Measuring \jpsi production near threshold is challenging due to the low cross section and small available phase space. The currently published world data on near-threshold \jpsi photoproduction come from two Jefferson Lab experiments: \jpsiexp in Hall C~\cite{Duran:2022xag} and GlueX in Hall D~\cite{Ali:2019lzf,GlueX:2023pev}. Both datasets have comparable statistics (about 2000 \jpsi events). \jpsiexp measured the first high-resolution two-dimensional cross section: differential in the Mandelstam variable $t$, and in ten photon energy slices between 9.1 and 10.6\,GeV. The GlueX collaboration  provided the $t$-dependent cross section in three bins in photon energy between 8.2 and 11.8\,GeV.

\looseness=-1
Two main theoretical approaches have been used to date to extract gluonic GFFs from near-threshold \jpsi data. In the holographic QCD framework~\cite{Mamo:2021krl, *Mamo:2022eui}, the photoproduction process is modeled via graviton and dilaton exchanges in Anti-de Sitter space, corresponding to tensor and scalar gluonic interactions. Alternatively, the generalized parton distribution (GPD) method~\cite{Guo:2023pqw} connects the photoproduction amplitude to gluonic GFFs through factorization at sufficiently large skewness $\xi$ (the longitudinal momentum asymmetry between initial and final states). Recent work has expanded this approach by introducing Bayesian fits to extract the gluonic GFFs \Agt and \Cgt from experimental data~\cite{Guo:2025jiz}. Meanwhile, lattice QCD predictions for these GFFs~\cite{Shanahan:2018pib,*Shanahan:2018nnv,Pefkou:2021fni} offer a theoretical benchmark for interpreting experimental results. 

\looseness=-1
In this work, we present new results using the muon (\jpsi~$\to \mu^+\mu^-$) decay channel of \jpsi photoproduction near threshold, doubling the available statistics from the \jpsiexp experiment. 
It has been suggested that the GlueX results from Ref.~\cite{GlueX:2023pev} show fluctuations hinting at open charm production~\cite{JointPhysicsAnalysisCenter:2023qgg}. Substantial contributions from open charm production could complicate the connection of near-threshold \jpsi production to the GFFs.
We extract the one-dimensional \jpsi cross section as a function of photon energy $E_\gamma$ from the combined \jpsiexp data to look for hints of open charm production.
Finally, we explore the impact of our combined data on our understanding of the mechanical structure of the proton.

\section{Experiment}
We conducted the \jpsiexp experiment~\cite{JLab_E12-16-007_2016} in 2019 in Hall C at Jefferson Lab using a high-intensity electron beam incident on a copper radiator (9\% radiation length) to produce bremsstrahlung photons that strike on a liquid hydrogen target.
The resulting exclusive \jpsi photoproduction events were detected using the High Momentum Spectrometer (HMS)
and the Super High Momentum Spectrometer (SHMS)~\cite{Ali:2025dan}.
The experiment included four kinematic settings with distinct spectrometer momenta and angles to provide a two-dimensional coverage in photon energy $E_\gamma$ and $t$ near the threshold. 
The trigger required a coincidence between the hodoscopes of both spectrometers.
More details on the experimental conditions can be found in Ref.~\cite{Duran:2022xag}.

While the Hall C spectrometers were not optimized for muon detection, the longitudinal segmentation of the electromagnetic calorimeters provides some sensitivity to separate minimum ionizing particles (MIPs) from showering hadrons. Furthermore, for part of our data, the Cherenkov detectors in both spectrometers are above the muon threshold, while below the pion threshold.
\new{We select \jpsi candidate events by requiring coincident MIP signals in both spectrometers.}

\begin{figure}[thb]
    \includegraphics[width=1.0\columnwidth,
                     trim={0 0 0 0},clip]{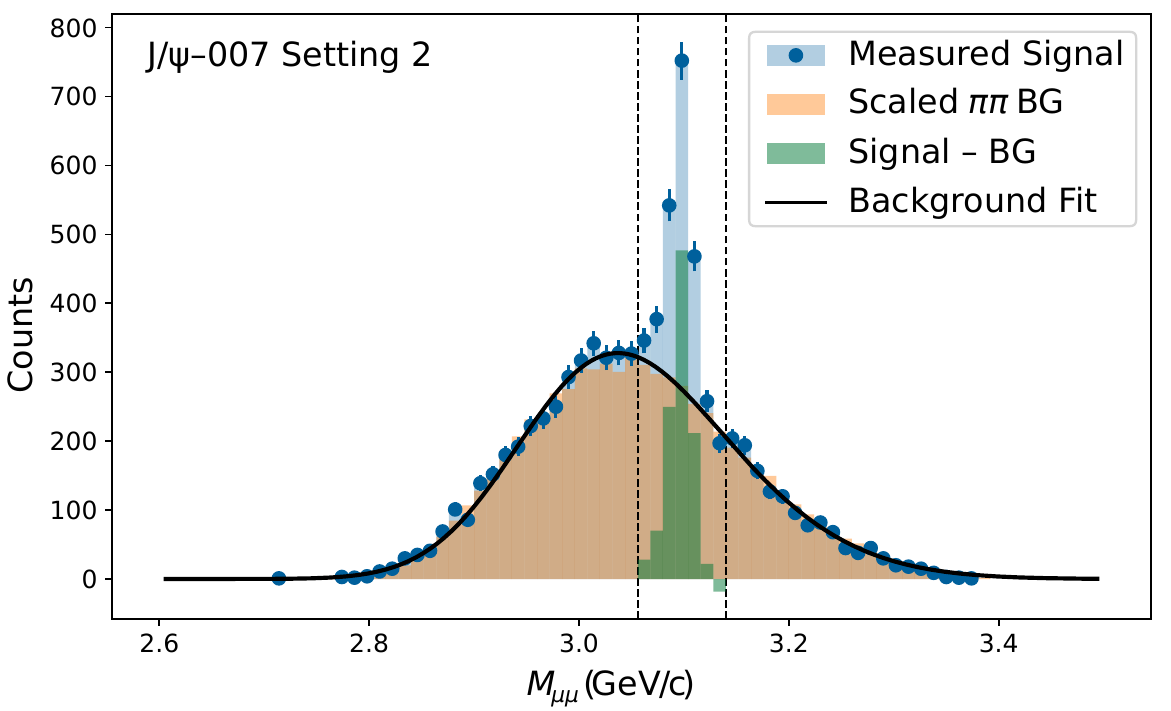}
    \caption{Di-MIP invariant mass spectrum for kinematic setting 2. 
    The figure shows the di-MIP signal sample (blue markers and histogram), di-pion background sample scaled to the signal sidebands (orange histogram), background fit (black curve), and background-subtracted di-muon signal (green histogram). The dashed lines indicate the integration window.}
    \label{fig:muon_invmass}
    \vspace*{-0.5cm}
\end{figure}

For the analysis in this paper we used the calorimeters for two-MIP event selection, while we used the Cherenkov detectors to benchmark the performance of the calorimeter-based event selection.
Fig.~\ref{fig:muon_invmass} shows the measured di-MIP invariant mass spectrum for experimental setting 2, with an estimated $1041\pm55$ \jpsi events after background subtraction.
Background events primarily originate from $\pi^+\pi^-$ pairs that behave as MIPs in both calorimeters. To subtract this background, we selected di-pion events where at least one of the pions showers in the calorimeter. We normalized this di-pion background to the sidebands in the \jpsi invariant mass spectrum. Then, for each bin, we fit the background spectrum with a skewed Gaussian shape to reduce statistical fluctuations inside the integration window.
As expected, the overall number of measured \jpsi events between the electron and muon channels is roughly the same.
The $\mu^+\mu^-$ analysis uses the same binning in $E_\gamma$ as the published $e^+e^-$ results, while we re-optimized the binning in $t$ for the di-pion background subtraction. 

Efficiencies, including trigger electronics, live-time corrections, tracking efficiency corrections, and particle identification (PID) efficiency corrections, were thoroughly studied and corrected for each experimental setting, analogous to the approach in Ref.~\cite{Duran:2022xag}. The PID efficiency for di-muon event detection using the calorimeter system was found to be $90.5\%\pm1.8\%$, determined using a clean di-muon sample selected with the Cherenkov counters. The detector effects and finite acceptance were corrected using an iterative unfolding procedure, similar to Ref.~\cite{Duran:2022xag}. The procedure was validated using detailed Monte Carlo simulations.
All Monte-Carlo event samples were generated using the lAger Monte Carlo generator~\cite{Joosten:2021lager}, and detector effects were simulated using the standard Hall C Monte Carlo program SIMC, modified to handle muons.
The integrated luminosity for each experimental setting was determined from the accumulated beam charge and the well-known density and length of the hydrogen target. Beam charge was monitored continuously using beam current monitors.

Systematic uncertainties on the measured cross sections consist of point-to-point and overall scale contributions. Point-to-point uncertainties primarily originate from background subtraction procedures, acceptance corrections, bin migration effects due to detector resolution, and statistical fluctuations in the normalization of the pion background sample. These uncertainties were evaluated independently for each kinematic bin, typically resulting in contributions on the order of a few percent, which are far smaller than the statistical uncertainties.

The scale uncertainty arises from the detector acceptance correction~(3\%), PID efficiency correction~(2\%), residual rate dependence correction~(1.2\%), luminosity measurement, vertex correction, target window subtraction, electroproduction subtraction, and radiator thickness~(each 1\%), and several corrections that contribute below the percent level.
Combining these contributions in quadrature, the total scale uncertainty for this measurement is estimated to be 4.6\%, comparable to but slightly larger than the 4\% scale uncertainty in Ref.~\cite{Duran:2022xag}.

\section{Results}
\begin{figure}[htb]
    \includegraphics[width=\columnwidth,
                     trim={0 0 0 0},clip]{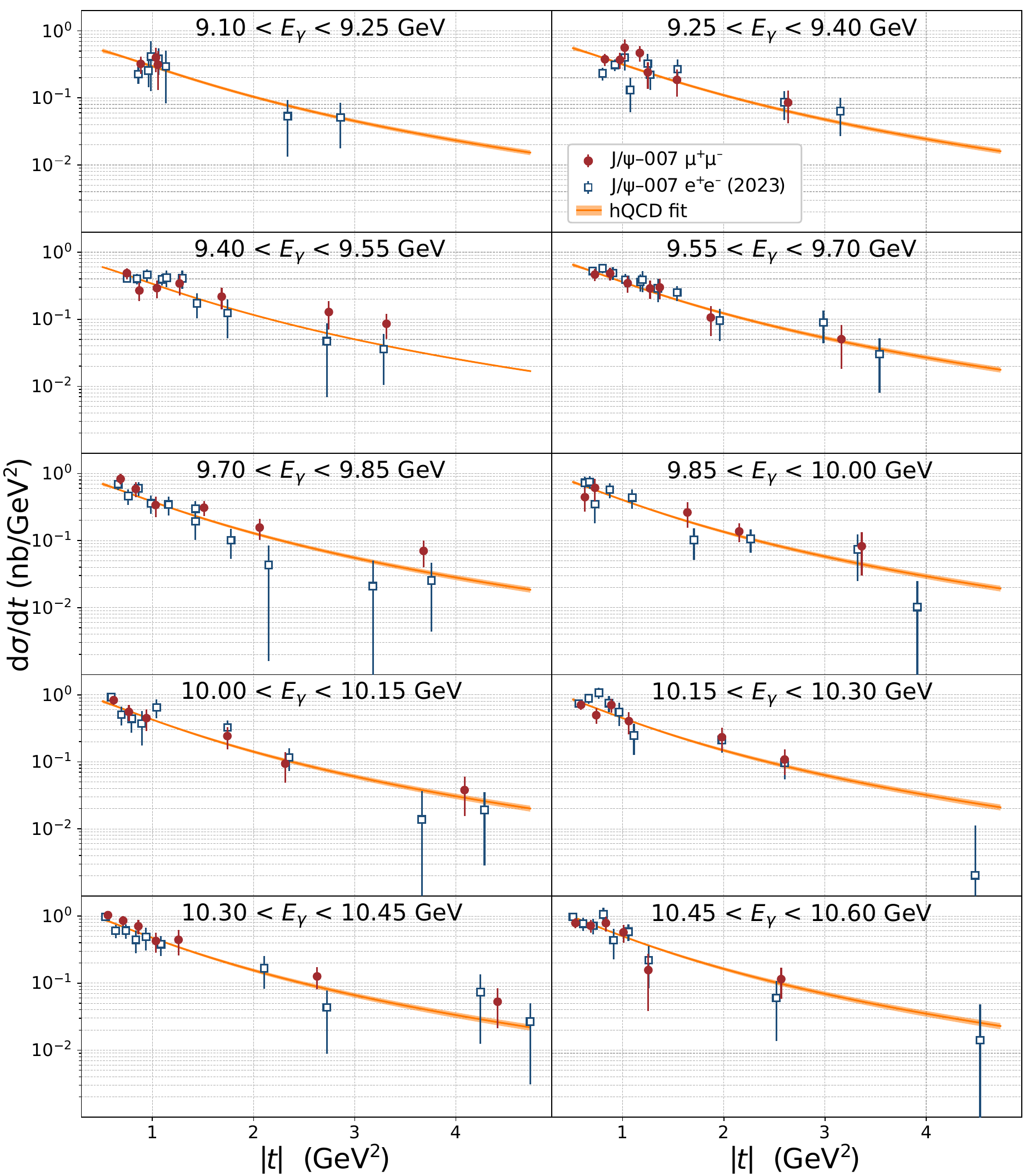}
    \caption{Differential cross section for \jpsi~$\to \mu^+\mu^-$ as a function of $|t|$ in slices of $E_\gamma$ (red circles) compared to the previous \jpsi~$\to e^+e^-$ results from Ref.~\cite{Duran:2022xag} (blue squares).  The orange band shows the 2D holographic QCD~\cite{Mamo:2021krl,*Mamo:2022eui} fit to the combined \jpsiexp data, with its associated uncertainty.}
    \label{fig:cross_section}
    \vspace*{-0.0cm}
\end{figure}

Fig.~\ref{fig:cross_section} shows our differential cross section measurements 
as a function of $|t|$ in slices of $E_\gamma$. The new muon-channel results and previous electron-channel results from Ref.~\cite{Duran:2022xag} agree within statistical uncertainties.

\begin{figure}[thb]
    \includegraphics[width=1.0\columnwidth,
    trim={0.0cm 0.2cm 0 0.2cm},clip]{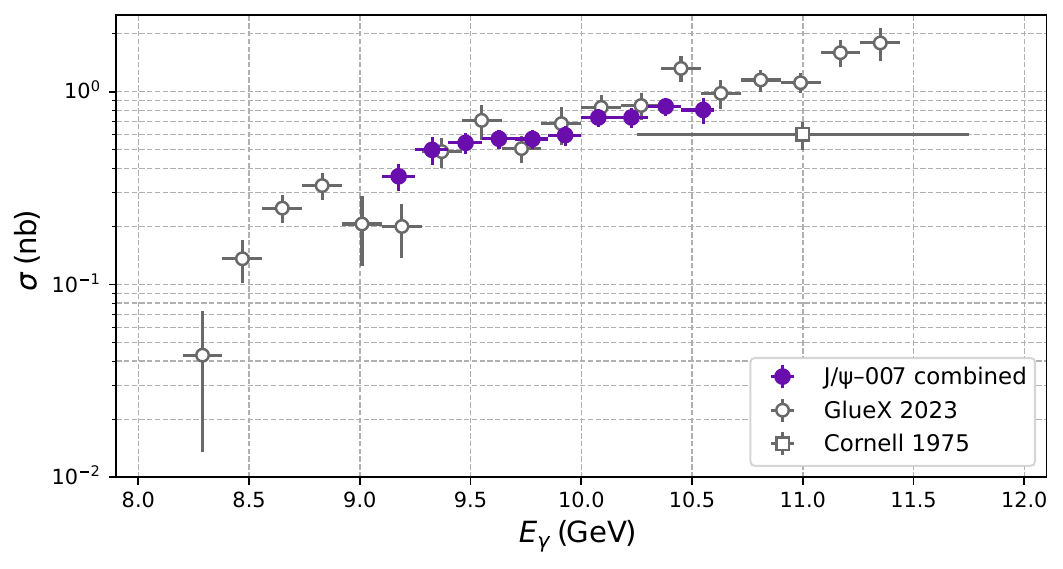}
    \caption{Integrated cross section for the combined \jpsiexp data as a function of $E_\gamma$, (purple squares), compared to GlueX~\cite{GlueX:2023pev} (open crosses) and Cornell~\cite{Camerini:1975cy} (open square).}
    \label{fig:cross_section_1d}
\end{figure}

Ref.~\cite{JointPhysicsAnalysisCenter:2023qgg} argued that the results from Ref.~\cite{GlueX:2023pev} show hints of fluctuations consistent with the onset of open-charm production in the near-threshold region. \new{In particular, the signature from a contribution from the $\bar{D}^*\Lambda_c$ channel was found to be a clear threshold cusp at about 9.5 GeV.
To look for such signatures,} we integrated the combined \jpsiexp data using a dipole fit. We repeated the integration using an exponential ansatz to estimate the model uncertainty. 
This model uncertainty is small, since differences appear mostly at high $t$. 
Fig.~\ref{fig:cross_section_1d} shows the integrated cross section results as a function of $E_\gamma$, compared with those from GlueX~\cite{GlueX:2023pev} and Cornell~\cite{Camerini:1975cy}.
While our integrated results are consistent with those from GlueX, \new{they do not exhibit signatures of cusps or fluctuations that would suggest sizable open-charm contributions.}
Our 1D and 2D cross section results are provided as supplemental material~\footnote{See Supplemental Material at [URL will be inserted by publisher] for the tabulated 1D and 2D cross section data.}.

\begin{figure}[htb]
    \includegraphics[width=\columnwidth,
                     trim={0.0cm 0.1cm 0 0.1cm},clip]{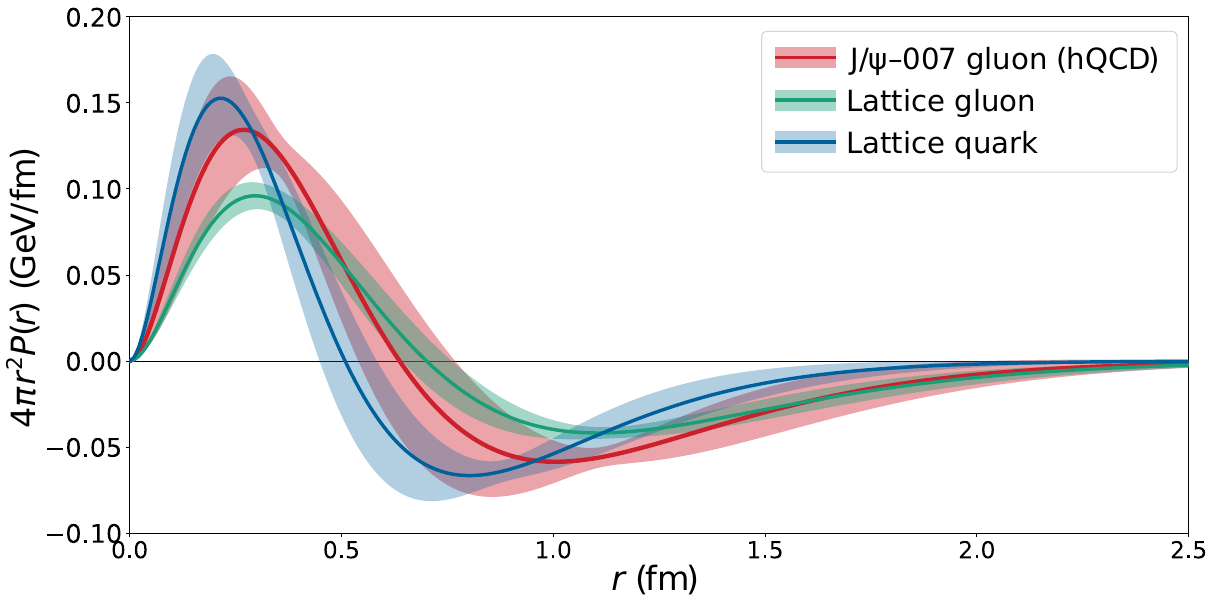}
    \vspace{-2em}
    \caption{Gluonic pressure distribution in the Breit frame in GeV/fm, extracted from our combined data using \new{a holographic QCD approach}~\cite{Mamo:2021krl,*Mamo:2022eui}~(red), compared with gluon (green) and quark (blue) results from lattice QCD~\cite{Pefkou:2021fni}, based on a dipole-tripole GFF fit.}
    \label{fig:pressure}
    \vspace*{-2em}
\end{figure}

Finally, we used our combined results to extract the gluonic GFFs, similar to the approach in Ref.~\cite{Duran:2022xag}.
The curve in Fig.~\ref{fig:cross_section} corresponds to a simultaneous multidimensional fit of the combined electron- and muon-channel results,
following the holographic QCD approach of Ref.~\cite{Mamo:2021krl,*Mamo:2022eui}, using a dipole shape for \Agt and a tripole shape for \Cgt, motivated by Ref.~\cite{Sun:2021gmi,*Tong:2022zax,*Tanaka:2018wea}. The fit yields a $\chi^2$ per degree of freedom close to unity.
We also attempted to follow the GPD approach of Ref.~\cite{Guo:2023pqw}, based on the suggestion from Ref.~\cite{Guo:2025jiz} that $\xi>0.5$ could be sufficient for the formalism to be valid.
However, we found that restricting the GPD fit to $\xi > 0.5$ leads to a poorly constrained $\chi^2$ surface due to a lack of precise measurements at higher values of $t$.
The doubled statistics of the combined data set enabled us to relate the extracted \Cgt GFF to the gluonic pressure and shear distributions in the Breit frame, following the approach of Ref.~\cite{Lorce:2018egm}, while neglecting contributions from \Cbargt.
\new{We note that the interpretation of this quantity as a pressure is not without controversy, with some recent works arguing for an interpretation in terms of momentum current densities instead~\cite{Ji:2025gsq,* Ji:2025qax}.}
The results from our extraction are shown in Fig.~\ref{fig:pressure}, compared with recent lattice QCD predictions for gluons and quarks~\cite{Pefkou:2021fni} 
using a new tripole fit to the \Cqgt GFFs~\footnote{Graciously provided by the authors of Ref.~\cite{Pefkou:2021fni} to allow for a direct comparison  with our experimental extraction.}.
The results from our holographic QCD extraction agree within uncertainties with the gluon results from lattice QCD.
Comparing the gluon results with the quark results from lattice, \new{we observe that while quarks dominate at the smallest radii, gluons dominate at larger radii with a confining inward pressure.} 
This overall conclusion \new{remains} valid when considering the experimental results on the quark pressure from Ref.~\cite{Burkert:2018bqq}.
While we ignored the contribution of \Cbargt in this extraction, nonzero values of \Cbargt = -\Cbarqt would further enhance our \new{finding by shifting the quark pressure up to larger positive values while shifting the gluon pressure down to more negative values.}

\section{Conclusion}
We have presented the first measurement of near-threshold \jpsi~$\to \mu^+\mu^-$ photoproduction with the {\jpsiexp} experiment at Jefferson Lab, \new{adding to our earlier electron-channel results and doubling the available statistics}. 
After integration over $t$, our data show no evidence for open-charm contributions near threshold.
A combined analysis of both decay channels following holographic QCD provides improved experimental constraints on the gluonic GFF \Cgt, sufficient to extract derived quantities such as the gluonic pressure. \new{Our} results agree with recent lattice QCD calculations. Taken together with quark pressure distributions from lattice QCD or DVCS measurements, our results support a spatial picture \new{in which gluons dominate at larger radii with a confining inward pressure.}
\new{Future high-statistics measurements with} SoLID--\jpsi at Jefferson Lab and at the Electron-Ion Collider, will be essential to reduce model uncertainties in GFF extractions and to achieve a precise understanding of the proton's mechanical structure.

\begin{acknowledgments}
We thank Dimitra Pefkou, Ismail Zahed, and Kiminad Mamo for valuable discussions and support.
This work was supported by the US DOE Office of Science, Nuclear Physics, under contract numbers DE-AC02-06CH11357 and DE-FG02-94ER40844, and contract number DE-AC05-06OR23177, under which JSA operates TJNAF.
\end{acknowledgments}
\bibliography{manuscript}

\end{document}

%% file: authors.tex
\author{S.~Joosten}
\email[Corresponding author: ]{sjoosten@anl.gov}
\affiliation{Argonne National Laboratory, Lemont, IL 60439, USA}

\author{Z.-E.~Meziani}
\affiliation{Argonne National Laboratory, Lemont, IL 60439, USA}

\author{S.~Prasad}
\affiliation{Argonne National Laboratory, Lemont, IL 60439, USA}

\author{J.~Swartz}
\affiliation{Argonne National Laboratory, Lemont, IL 60439, USA}
\affiliation{University of Chicago, Chicago, IL 60637, USA}

\author{B.~Duran}
\altaffiliation{Now at New Mexico State University}
\affiliation{Temple University, Philadelphia, PA 19122, USA}

\author{M.~K.~Jones}
\affiliation{Thomas Jefferson National Accelerator Facility, Newport News, VA 23606, USA}

\author{H.~Klest}
\affiliation{Argonne National Laboratory, Lemont, IL 60439, USA}

\author{M.~Paolone}
\altaffiliation{Now at New Mexico State University}
\affiliation{Temple University, Philadelphia, PA 19122, USA}

\author{C.~Peng}
\affiliation{Argonne National Laboratory, Lemont, IL 60439, USA}

\author{W.~Armstrong}
\affiliation{Argonne National Laboratory, Lemont, IL 60439, USA}

\author{H.~Atac}
\affiliation{Temple University, Philadelphia, PA 19122, USA}

\author{E.~Chudakov}
\affiliation{Thomas Jefferson National Accelerator Facility, Newport News, VA 23606, USA}

\author{H.~Bhatt}
\affiliation{Mississippi State University, Mississippi State, MS 39762, USA}

\author{D.~Bhetuwal}
\affiliation{Mississippi State University, Mississippi State, MS 39762, USA}

\author{M.~Boer}
\affiliation{Virginia Polytechnic Institute \& State University, Blacksburg, VA 24061, USA}

\author{A.~Camsonne}
\affiliation{Thomas Jefferson National Accelerator Facility, Newport News, VA 23606, USA}

\author{J.-P.~Chen}
\affiliation{Thomas Jefferson National Accelerator Facility, Newport News, VA 23606, USA}

\author{M.~Dalton}
\affiliation{Thomas Jefferson National Accelerator Facility, Newport News, VA 23606, USA}

\author{N.~Deokar}
\affiliation{Temple University, Philadelphia, PA 19122, USA}

\author{M.~Diefenthaler}
\affiliation{Thomas Jefferson National Accelerator Facility, Newport News, VA 23606, USA}

\author{J.~Dunne}
\affiliation{Mississippi State University, Mississippi State, MS 39762, USA}

\author{L.~El Fassi}
\affiliation{Mississippi State University, Mississippi State, MS 39762, USA}

\author{F.~Flor}
\affiliation{Argonne National Laboratory, Lemont, IL 60439, USA}

\author{E.~Fuchey}
\affiliation{University of Connecticut, Storrs, CT 06269, USA}

\author{H.~Gao}
\affiliation{Duke University, Durham, NC 27708, USA}

\author{D.~Gaskell}
\affiliation{Thomas Jefferson National Accelerator Facility, Newport News, VA 23606, USA}

\author{O.~Hansen}
\affiliation{Thomas Jefferson National Accelerator Facility, Newport News, VA 23606, USA}

\author{F.~Hauenstein}
\affiliation{Old Dominion University, Norfolk, VA 23529, USA}

\author{D.~Higinbotham}
\affiliation{Thomas Jefferson National Accelerator Facility, Newport News, VA 23606, USA}

\author{S.~Jia}
\affiliation{Temple University, Philadelphia, PA 19122, USA}

\author{A.~Karki}
\affiliation{Mississippi State University, Mississippi State, MS 39762, USA}

\author{C.~Keppel}
\affiliation{Thomas Jefferson National Accelerator Facility, Newport News, VA 23606, USA}

\author{P.~King}
\affiliation{Ohio University, Athens, OH 45701, USA}

\author{H.S.~Ko}
\affiliation{Université Paris-Saclay, Gif-sur-Yvette, 91190 Essonne, France}

\author{X.~Li}
\affiliation{Duke University, Durham, NC 27708, USA}

\author{R.~Li}
\affiliation{Temple University, Philadelphia, PA 19122, USA}

\author{D.~Mack}
\affiliation{Thomas Jefferson National Accelerator Facility, Newport News, VA 23606, USA}

\author{S.~Malace}
\affiliation{Thomas Jefferson National Accelerator Facility, Newport News, VA 23606, USA}

\author{M.~McCaughan}
\affiliation{Thomas Jefferson National Accelerator Facility, Newport News, VA 23606, USA}

\author{R.~E. McClellan}
\affiliation{Pensacola State College, Pensacola, FL 32504, USA}

\author{R.~Michaels}
\affiliation{Thomas Jefferson National Accelerator Facility, Newport News, VA 23606, USA}

\author{D.~Meekins}
\affiliation{Thomas Jefferson National Accelerator Facility, Newport News, VA 23606, USA}

\author{L.~Pentchev}
\affiliation{Thomas Jefferson National Accelerator Facility, Newport News, VA 23606, USA}

\author{E.~Pooser}
\affiliation{Thomas Jefferson National Accelerator Facility, Newport News, VA 23606, USA}

\author{A.~Puckett}
\affiliation{University of Connecticut, Storrs, CT 06269, USA}

\author{R.~Radloff}
\affiliation{Ohio University, Athens, OH 45701, USA}

\author{M.~Rehfuss}
\affiliation{Temple University, Philadelphia, PA 19122, USA}

\author{P.~E. Reimer}
\affiliation{Argonne National Laboratory, Lemont, IL 60439, USA}

\author{S.~Riordan}
\affiliation{Argonne National Laboratory, Lemont, IL 60439, USA}

\author{B.~Sawatzky}
\affiliation{Thomas Jefferson National Accelerator Facility, Newport News, VA 23606, USA}

\author{A.~Smith}
\affiliation{Duke University, Durham, NC 27708, USA}

\author{N.~Sparveris}
\affiliation{Temple University, Philadelphia, PA 19122, USA}

\author{H.~Szumila-Vance}
\affiliation{Thomas Jefferson National Accelerator Facility, Newport News, VA 23606, USA}

\author{S.~Wood}
\affiliation{Thomas Jefferson National Accelerator Facility, Newport News, VA 23606, USA}

\author{J.~Xie}
\affiliation{Argonne National Laboratory, Lemont, IL 60439, USA}

\author{Z.~Ye}
\affiliation{Argonne National Laboratory, Lemont, IL 60439, USA}

\author{C.~Yero}
\affiliation{Old Dominion University, Norfolk, VA 23529, USA}

\author{Z.~Zhao}
\affiliation{Duke University, Durham, NC 27708, USA}

\collaboration{The J/$\psi$--007 Collaboration}
\noaffiliation{}